\documentclass[pra,twocolumn,showpacs,preprintnumbers,amsmath,amssymb]{revtex4-1}

\usepackage{graphicx}
\usepackage{dcolumn}
\usepackage{bm}
\usepackage{comment}
\usepackage{bbold}

\begin{document}

\title{Reply to Comment on ``Anomalous Edge State in a Non-Hermitian Lattice''}
\date{\today}

\author{Tony E. Lee}
\affiliation{Department of Physics, Indiana University Purdue University Indianapolis (IUPUI), Indianapolis, Indiana 46202, USA}

\begin{abstract}
The Comment by Xiong \textit{et al.}~(arXiv:1610.06275) criticizing my Letter [Phys.~Rev.~Lett.~116, 133903 (2016)] was rejected by Physical Review Letters. In this Reply, I show that all their claims are wrong.
\end{abstract}

\maketitle

Xiong \textit{et al.}~posted a Comment \cite{xiong16} that was rejected by Physical Review Letters. They dispute the conclusions of my Letter \cite{lee16a}. They make three claims: (i) the winding number, calculated using Berry phase, is not 1/2, (ii) the real spectrum of an open chain is a finite-size effect, and (iii) the breakdown of the bulk-boundary correspondence is not due to the defectiveness of the Hamiltonian. In this Reply, I show that they are wrong on every point.


\section{Winding number}

In my Letter \cite{lee16a}, I said that the winding number has a fractional value of 1/2 because $k$ must sweep through $4\pi$ in order for $(\langle\sigma_x\rangle, \langle\sigma_z\rangle)$ to make one complete loop around the origin. Xiong \textit{et al.}~claim that according to the Berry phase, the winding number is not 1/2. Here, I show that the Berry phase does in fact verify that the winding number is 1/2.

Consider the momentum-space Hamiltonian $H_k$ given in Eq.~(1) of the Letter \cite{lee16a}. When $H_k$ encircles an exceptional point, one needs to be careful when calculating the winding number, since the eigenvectors are $4\pi$-periodic in $k$. The proper expression for the winding number in this case is 
\begin{eqnarray}
w&=&\frac{1}{2\pi}\gamma_B, \label{eq:w_me}
\end{eqnarray}
where $\gamma_B$ is the Berry phase when $k$ increases by $4\pi$:
\begin{eqnarray}
\gamma_B&=&i\int_0^{4\pi} dk \frac{\langle\langle u|\partial_k|u \rangle}{\langle\langle u|u \rangle}, \label{eq:gamma_me}
\end{eqnarray}
where $|u\rangle\rangle$ and $|u\rangle$ are left and right eigenvectors of $H_k$. It is important to integrate over $4\pi$ in $k$ so that the eigenvectors return to their original values (see below). To account for this, there is a factor of $1/2$ in Eq.~\eqref{eq:w_me} because the Brillouin zone is swept through twice.

In Ref.~\cite{keck03}, it was shown analytically that $\gamma_B=\pi$ when $H_k$ circles an exceptional point. Thus, the winding number is $w=1/2$. In other words, one must sweep through the Brillouin zone \textit{twice} in order to pick up a Berry phase of $\pi$.

Xiong \textit{et al.}~used an unphysical expression for the winding number:
\begin{eqnarray}
w&=&\frac{1}{\pi}\tilde{\gamma}_B, \label{eq:w_xiong}
\end{eqnarray}
where $\tilde{\gamma}_B$ is the Berry phase when $k$ increases by $2\pi$:
\begin{eqnarray}
\tilde{\gamma}_{B}&=& i\int_0^{2\pi} dk \frac{\langle\langle u|\partial_k|u \rangle}{\langle\langle u|u \rangle}. \label{eq:gamma_xiong}
\end{eqnarray}
It is well known that a Berry phase is physically meaningful only if the eigenvector makes a closed loop during the integration; otherwise, the Berry phase will be gauge dependent. When circling an exceptional point, the eigenvectors are $4\pi$-periodic in $k$. Since Xiong \textit{et al.}~in Eq.~\eqref{eq:gamma_xiong} integrated over only $2\pi$ in $k$, it is no surprise that they find a gauge-dependent Berry phase and winding number. To get physically meaningful results, one needs to integrate over $4\pi$ in $k$ so that the eigenvectors make a closed loop. That is what I did in Eq.~\eqref{eq:gamma_me}.


Xiong \textit{et al.}'s argument about the ``demo model'' is nonsense. Their Eq.~(A6) is wrong and should be replaced with Eqs.~\eqref{eq:w_me} and \eqref{eq:gamma_me} (using the usual inner product). If one decides to double the Brillouin zone to $4\pi$, one should integrate over $4\pi$ in $k$. Furthermore, $A=2$ since the Brillouin zone is swept through twice. Then one finds a winding number of 1, as expected. The important point is: \emph{for Hermitian systems, Eqs.~\eqref{eq:w_me} and \eqref{eq:w_xiong} give the same result, but when an exceptional point is involved, one needs to use Eq.~\eqref{eq:w_me} to get physically meaningful results.}

\section{Real spectrum}

In my Letter \cite{lee16a}, I said that a chain with open boundary conditions has real eigenvalues for $|v|\geq \gamma/2$. Xiong \textit{et al.}~claim that this is a finite-size effect \cite{xiong16}. As evidence, they present  their Fig.~1(e), which supposedly shows complex eigenvalues for large $N$. However, their result is an artifact due to flawed numerics.

As discussed in my Letter \cite{lee16a}, the Hamiltonian $H$ has two symmetries: chiral symmetry and $\mathcal{PT}$ symmetry. Due to chiral symmetry, if $E$ is an eigenvalue of $H$, $-E$ is also an eigenvalue. Due to $\mathcal{PT}$ symmetry, if $E$ is an eigenvalue of $H$, $E^*$ is also an eigenvalue. As a result, when the eigenvalues are plotted in the complex plane, they should be symmetric around both the real and imaginary axes. This is a simple sanity check.

\begin{figure}[t]
\centering
\includegraphics[width=3.3 in,trim=1.6in 3.4in 1.7in 3.4in,clip]{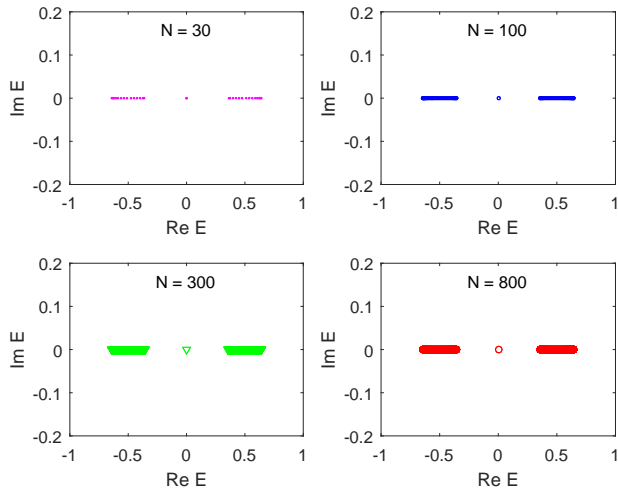}
\caption{\label{fig:eigenvalues_N}Eigenvalues for chains with open boundary conditions and different sizes $N$. The parameters are the same as Fig.~1(e) of Ref.~\cite{xiong16}: $\gamma=1$, $r=0.5$, $v=0.52$.}
\end{figure}

The eigenvalues calculated by Xiong \textit{et al.}~in their Fig.~1(e) do not obey these symmetries \cite{xiong16}: their eigenvalues are not symmetric around the real or imaginary axes. Thus, their calculation is flawed, and their observation of complex eigenvalues is nonsense. I do not know how they screwed up, but when I (correctly) calculate the eigenvalues, the eigenvalues are purely real, even for large $N$. The correct eigenvalues are shown in Fig.~\ref{fig:eigenvalues_N}.

\section{Defectiveness}

In my Letter \cite{lee16a}, I said that the bulk-boundary correspondence is broken because $H$ is defective, such that there is only one $E=0$ edge state, as opposed to the even number of edge states found in typical topological insulators. Xiong \textit{et al.}~claim that the breakdown of the bulk-boundary correspondence is not due to the defectiveness of $H$ \cite{xiong16}. But my Letter showed analytically that the $E=0$ eigenvalue is defective, leading to only one $E=0$ edge state. This by itself shows the breakdown of the bulk-boundary correspondence.

Xiong \textit{et al.}~claim that the breakdown is instead due to the ``bulk'' (in-gap) states being localized on the edge \cite{xiong16}. But the in-gap states are localized on the edge precisely because $H$ is defective. As mentioned in the Letter, when $v=\gamma/2$, the eigenvalues at $E=\pm r$ (corresponding to in-gap states) are highly defective \cite{lee16a}. One can show analytically that the corresponding eigenvectors are 
\begin{eqnarray}
(\pm 1 + \frac{i\gamma}{r},\pm i+\frac{\gamma}{r},i,1,0,0,\ldots)^T, \label{eq:eigenvec_er}
\end{eqnarray}
so all the in-gap states are localized on the left edge. What is amusing is that even if Xiong \textit{et al.}~are right about the in-gap states, the breakdown is still due to the defectiveness of $H$.



\bibliography{edge}

\end{document}